\documentclass[journal]{IEEEtran}

\usepackage{epsfig,rotating,setspace,latexsym,amsmath,epsf,amssymb,amsfonts,bm,theorem,cite,authblk, bbm,color, hyperref, multirow, caption, graphicx, subcaption, url, xcolor}

\IEEEoverridecommandlockouts
\allowdisplaybreaks

\title{AI-Native Task-Oriented Semantic Communications over a Relay: Benefits and Vulnerabilities}

\author[1]{Yalin E. Sagduyu}
\author[1]{Tugba Erpek}
\author[2]{Aylin Yener}

\affil[1]{\normalsize  Nexcepta, Gaithersburg, MD, USA}

\affil[2]{\normalsize  The Ohio State University, Columbus, OH, USA}

\begin{document}
\bstctlcite{BSTcontrol}

\pagestyle{empty}
\maketitle
\thispagestyle{empty}

\begin{abstract}
Semantic communication (SemCom) using a relay node is considered, where the relay assists in conveying task-relevant information from a source to a destination, with meaning defined by classification labels. A joint end-to-end design of the source encoder, relay transformation, and destination decoders is developed to optimize both semantic inference accuracy and perceptual reconstruction quality for image data. Relay-assisted schemes are shown to outperform direct transmission, with gains that become more pronounced in low signal-to-noise ratio (SNR) regimes across different relaying strategies. Alongside these performance benefits, however, a key security vulnerability introduced by the relay is identified, and a Byzantine attack is formulated in which the relay injects structured, learned perturbations into its transmitted signal. The attack operates on intermediate communication representations subject to channel effects, with perturbations generated through a learned mapping applied at test time. It is shown to significantly degrade semantic inference accuracy at the destination. These results reveal a clear trade-off between SemCom gains and security risks in relay-assisted systems, motivating the need for secure relay designs that account for insider threats.
\end{abstract}

\section{Introduction}

\emph{Semantic communication} (SemCom) departs from conventional systems by focusing on transmitting task-relevant information rather than exact data reconstruction \cite{guler2014, xie2021deep, gunduz2022beyond}. SemCom integrates learning-based representations into system design, enabling more efficient use of limited wireless resources, particularly in bandwidth- and power-constrained environments. By prioritizing semantic meaning over bit-level fidelity, SemCom improves robustness and efficiency for downstream tasks such as classification and perception.

While prior work on SemCom has largely focused on direct source-destination transmission, practical wireless systems often operate under challenging channel conditions where direct links are weak or unreliable. This motivates the incorporation of \emph{relay} nodes into SemCom systems \cite{luo2022autoencoder, lin2024semantic, guo2024distributed, arda2024semantic, sagduyu2026semantic}. In this paper, we consider relay-assisted SemCom, where the relay processes received signals using a Learn-and-Forward (LF) mechanism before forwarding them to the destination. The LF operation is jointly optimized with the source encoder and destination decoders, enabling diversity gains and semantic-aware processing to mitigate channel impairments. For comparison, we consider Amplify-and-Forward (AF) and Decode-and-Forward (DF) as baseline relay mechanisms operating with learned source encoders and destination decoders. As we will see in the sequel, and as noted in \cite{arda2024semantic} for a slightly different system model, relay-assisted schemes achieve higher semantic accuracy than direct transmission, with larger gains in lower signal-to-noise ratio (SNR) regimes.

These semantic performance gains, however, come with new \emph{security vulnerabilities}. Since SemCom often relies on deep neural networks (DNNs), it is inherently susceptible to manipulation during both training and inference \cite{sagduyu2023task, sagduyu2023semantic, sagduyu2023vulnerabilities, sagduyu2024will, sagduyu2025securing, sagduyu2026perturb}. In particular, the relay can itself be an \emph{adversary}. We present such a \emph{Byzantine attack} in which the relay injects structured perturbations into its transmitted signals. This attack proves to be especially effective in SemCom, where such perturbations in the latent representation can lead to significant degradation in semantic inference at the destination. Unlike conventional input-space gradient-based attacks that directly perturb the classifier input, the proposed attack operates on intermediate communication signals at the relay. It generates structured perturbations through a learned mapping, without requiring ground-truth labels or gradient-based optimization at test time.

The adversarial perturbation injected by the Byzantine relay is generated by a DNN operating on the relay's received signal. The DNN is trained with the communication model held fixed to degrade semantic inference at the destination by increasing prediction uncertainty while constraining perturbation magnitude. The learned mapping is applied at test time without modifying the communication model, enabling single-pass execution at the relay. We show that this attack is highly effective in reducing semantic accuracy across SNR regimes. This highlights the increased security risks introduced by semantic relaying despite its performance gains, motivating the need for robust SemCom designs.

The remainder of the paper is organized as follows. Sec.~\ref{sec:sem} presents the relay-assisted SemCom approach and evaluates its benefits. Sec.~\ref{sec:attack} introduces the insider threat model at the relay and demonstrates its impact on the performance of semantic relay communications. Sec.~\ref{sec:conclusion} concludes the paper.

\section{Semantic Relay Communications \label{sec:sem}}

\begin{figure}[t!]
  \centering
    \includegraphics[width=\linewidth]{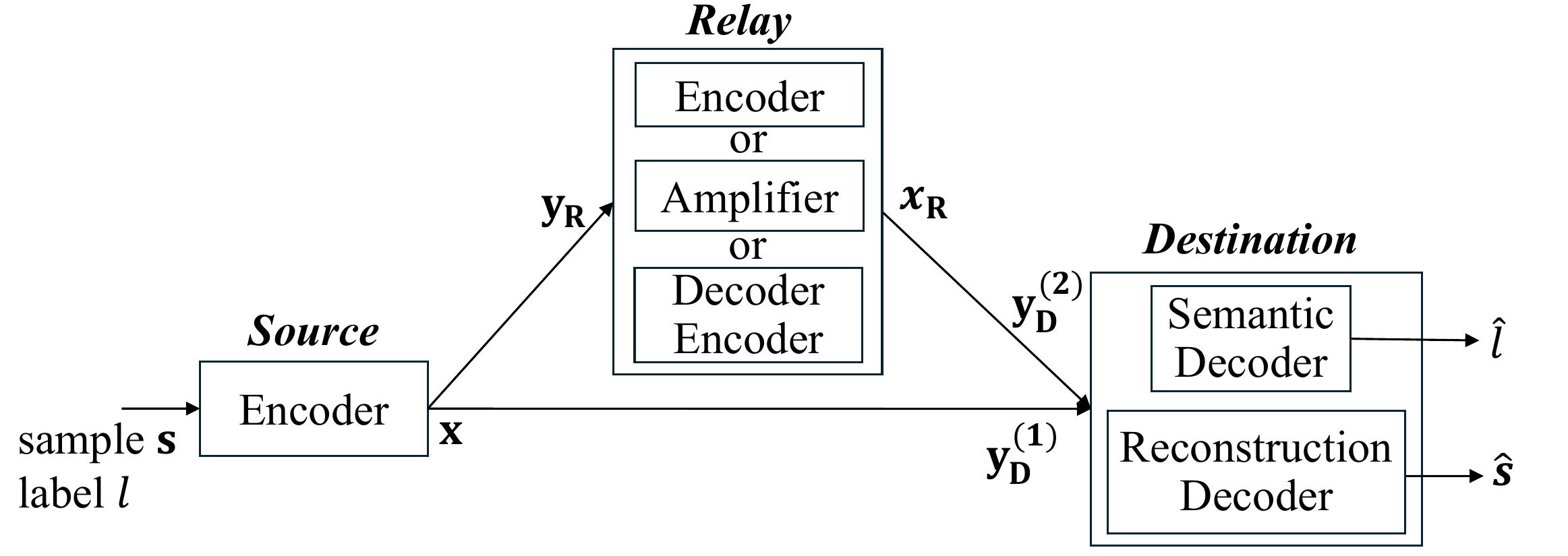}
    \caption{System model of semantic relay communications.}\label{fig:semrelaycom}
\end{figure}

We consider a network of three nodes (shown in Fig.~\ref{fig:semrelaycom}): a source node, a relay node, and a destination node. The source observes an input sample $\mathbf{s} \in \mathbb{R}^{C \times N_h \times N_w}$, where $C$ denotes the number of channels and $N_h$ and $N_w$ denote the spatial height and width of the input, respectively. For CIFAR-10 color images, $C\!=3\!$ and $N_h\!=\!N_w\!=\!32$. The objective is to transmit a compact representation of $\mathbf{s}$ such that the destination can both reconstruct the input and infer its label. The input sample is associated with a label $\ell \in \{1, \ldots, K\}$, where $K$ denotes the number of classes. The pair $(\mathbf{s}, \ell)$ represents the semantic information to be conveyed to the destination. The source encoder, parameterized by $\theta$, maps the input $\mathbf{s}$ into a transmitted representation $\mathbf{x} \in \mathbb{R}^{2 \times n}$ according to
\begin{equation}
\mathbf{x} = f^{S}_{\theta}(\mathbf{s}),
\end{equation}
where $n$ denotes the number of channel uses (the latent dimension). The two rows correspond to the in-phase and quadrature (I/Q) components, respectively. The encoder is a DNN with two convolutional blocks. Each block includes two convolutional layers with kernel size $3 \times 3$, followed by batch normalization and ReLU activations, and a $2 \times 2$ max-pooling layer for spatial downsampling. Dropout with probability $0.25$ is applied after each block to regularize training. The convolutional channel dimensions are
$C\!\rightarrow\!64\!\rightarrow\!64\!\rightarrow\!128\!\rightarrow\!128$. The resulting feature map is flattened and passed through two fully connected layers: the first maps to a hidden dimension of size 1024 with ReLU activation, and the second maps to a vector of dimension $2n$, which is reshaped into the $2 \times n$ transmitted representation. For $n=16$, the source encoder has 8.68M parameters. Prior to transmission, each source or relay output $\mathbf{z}$ is
power-normalized as $\mathcal{N}(\mathbf{z})$, where $\mathcal{N}(\cdot)$
denotes per-sample normalization that enforces unit average squared magnitude per real I/Q component.

The transmitted representation can be viewed as a complex vector of length $n$, namely $\mathbf{x} \in \mathbb{C}^{n}$, whose real and imaginary parts correspond to the I/Q components, respectively. 
In DNN operations, any complex-valued signal of dimension $n$ is represented via its I/Q components as a real-valued tensor of size $2 \times n$ for processing by the DNNs. 

We consider a two-phase transmission scheme. In the first phase, the source broadcasts $\mathbf{x}$ to the relay and destination. The received signal at the relay is
\begin{equation}
\mathbf{y}_{\text{R}} = h_{\text{SR}} \mathbf{x} + \mathbf{n}_{\text{R}},
\end{equation}
and the received signal at the destination in the first phase is
\begin{equation}
\mathbf{y}_{\text{D}}^{(1)} = h_{\text{SD}} \mathbf{x} + \mathbf{n}_{\text{D}}^{(1)},
\end{equation}
where $h_{\text{SR}}, h_{\text{SD}} \in \mathbb{C}$ denote flat-fading channel coefficients for the source-to-relay and source-to-destination links, respectively. The noise terms $\mathbf{n}_{\text{R}}, \mathbf{n}_{\text{D}}^{(1)} \in \mathbb{C}^{n}$ are additive circularly symmetric complex Gaussians. The wireless channels follow a block-fading model, where channel coefficients remain constant over the $n$ channel uses of each sample and vary independently across samples. The coefficients $h_{\mathrm{SR}}$ and $h_{\mathrm{SD}}$ follow Rayleigh fading. With the transmitted signals normalized, the noise variance is set according to the specified link SNRs. The relay and destination operate directly on the received I/Q samples without explicit channel state information.

In the second phase, which is orthogonal to the first phase (for example, using a separate time slot or frequency band), the relay processes its received signal $\mathbf{y}_{\mathrm{R}}$ using a relay transformation
$f^{R}_{\phi}(\cdot)$, where $\phi$ denotes trainable parameters when applicable, and transmits
\begin{equation}
\label{eq:general_relay_function}
\mathbf{x}_{\mathrm{R}}
=
f^{R}_{\phi}(\mathbf{y}_{\mathrm{R}}).
\end{equation}

The transmitted signal is normalized to satisfy the same average power constraint as the source. This relay mapping $f^{R}_{\phi}(\cdot)$ provides a common abstraction that captures different relaying strategies as special cases, including LF, AF, and DF. In all cases, the relay function (whether learned in LF and DF or fixed in AF) is incorporated into the end-to-end training of the source and destination DNNs.

\emph{(i) Amplify-and-forward (AF):} Using a fixed mapping with no trainable parameters, the relay transmits a normalized version of its received signal:
\begin{equation}
\mathbf{x}_{\text{R}} = \mathcal{N}\!\left(\mathbf{y}_{\text{R}}\right).
\end{equation}

\emph{(ii) Decode-and-forward (DF):} The relay mapping consists of a relay-side decoder and a relay-side encoder. The relay first maps its received latent signal to the original image dimensions using a relay-side decoder parameterized by $\phi_{\text{dec}}$,
\begin{equation}
\hat{\mathbf{s}}_{\text{R}} = f^{R,\text{dec}}_{\phi_{\text{dec}}}(\mathbf{y}_{\text{R}}),
\end{equation}
where $\hat{\mathbf{s}}_{\text{R}} \in \mathbb{R}^{C \times N_h \times N_w}$ denotes the resulting image-domain representation. The relay then re-encodes this estimate into a transmitted latent representation using a relay-side encoder parameterized by $\phi_{\text{enc}}$,
\begin{equation}
\mathbf{x}_{\text{R}} = f^{R,\text{enc}}_{\phi_{\text{enc}}}(\hat{\mathbf{s}}_{\text{R}}),
\end{equation}
where $\mathbf{x}_{\text{R}} \in \mathbb{C}^{n}$ denotes the relay transmit signal.

The relay-side decoder maps from the latent representation of dimension \(2\times n\) to the original image dimension \(C\times N_h\times N_w\), while the relay-side encoder maps from the image domain back to the latent representation of dimension \(2\times n\). In the numerical results, the relay-side decoder uses the same architecture as the reconstruction decoder at the destination, with its input dimension adjusted to the relay observation, while the relay-side encoder uses the same architecture as the source encoder.

\textit{(iii) Learn-and-forward (LF):} The relay mapping in (\ref{eq:general_relay_function}) is realized as a direct latent-to-latent transformation. The relay processes its received signal $\mathbf{y}_{\text{R}}$ and produces $\mathbf{x}_{\text{R}} = f^{R}_{\phi}(\mathbf{y}_{\text{R}})$,
where $f^{R}_{\phi}:\mathbb{C}^{n}\rightarrow \mathbb{C}^{n}$. Thus, $f^{R}_{\phi}(\cdot)$ in the LF case corresponds to a mapping from latent dimension to latent dimension. In DNN operations, the input is represented in its I/Q form. The mapping is implemented as a one-dimensional convolutional DNN operating across the latent dimension. Specifically, the input $\mathbf{y}_{\text{R}}$ is processed by an initial one-dimensional convolutional layer that maps the two input channels (I/Q) to a 128-channel hidden representation. This is followed by two residual blocks, each consisting of two convolutional layers with kernel size 3 and ReLU activations. Residual connections are used to stabilize training and enable effective denoising. A final convolutional layer maps the hidden representation back to two channels, producing the relay transmit signal. The output is then normalized to enforce a fixed average transmit power across channel uses.

Unlike DF, LF does not reconstruct the input in the image domain and instead learns a direct transformation in the latent space. For $n= 16$, DF employs 26.17M DNN parameters (17.49M at the decoder and 8.68M at the encoder), while the lightweight LF uses 199K parameters.

In each of the relay modes, the destination receives the relay transmission in the second phase as
\begin{equation}
\mathbf{y}_{\text{D}}^{(2)} = h_{\text{RD}} \mathbf{x}_{\text{R}} + \mathbf{n}_{\text{D}}^{(2)},
\end{equation}
where the relay-to-destination channel coefficient
$h_{\text{RD}} \in \mathbb{C}$ follows Rayleigh fading and
$\mathbf{n}_{\text{D}}^{(2)} \in \mathbb{C}^{n}$ is additive Gaussian noise. Since the two transmission phases are orthogonal, the overall signal received at the destination is:
\begin{equation}
\mathbf{y}_{\text{D}} = \left[ \mathbf{y}_{\text{D}}^{(1)}, \mathbf{y}_{\text{D}}^{(2)} \right] \in \mathbb{C}^{2n}.
\end{equation}
The concatenated representation $\mathbf{y}_{\text{D}}$ serves as the input to the semantic and reconstruction decoder DNNs at the destination. In DNN operations, $\mathbf{y}_{\text{D}}$ is represented in its I/Q form as a real-valued tensor of size $2 \times 2n$.

(i) The semantic decoder applies the mapping
\begin{equation}
\mathbf{p}(\mathbf{y}_{\text{D}})
=
f^{D,\text{cls}}_{\psi_{\text{cls}}}(\mathbf{y}_{\text{D}}),
\qquad
\hat{\ell}
=
\operatorname*{arg\,max}_{k\in\{1,\ldots,K\}}
p_k(\mathbf{y}_{\mathrm{D}}),
\end{equation}
parameterized by $\psi_{\text{cls}}$, where $\mathbf{p}(\mathbf{y}_{\text{D}})$ is the predicted class-probability
vector, $\hat{\ell}$ is the predicted class label, and $K$ is the number of classes. The decoder is a multilayer perceptron that takes the flattened latent representation as input and passes it through two fully connected layers of sizes 512 and 256, each followed by ReLU activation and dropout with probability 0.5. The final layer outputs class logits, which are converted to probabilities via the softmax function, yielding $\mathbf{p}(\mathbf{y}_{\text{D}})$. For $n=16$, the semantic decoder has 167K parameters.

(ii) The reconstruction decoder applies the mapping
\begin{equation}
\hat{\mathbf{s}} = f^{D,\text{rec}}_{\psi_{\text{rec}}}(\mathbf{y}_{\text{D}}),
\end{equation}
parameterized by $\psi_{\text{rec}}$, where $\hat{\mathbf{s}} \in \mathbb{R}^{C \times N_h \times N_w}$ denotes the reconstructed input. The decoder first flattens the input into a vector of dimension $4n$, passes it through two fully connected layers (mapping to 1024 and then to a feature tensor of size $256 \times \frac{N_h}{4} \times \frac{N_w}{4}$), and then applies two transposed convolutional layers with channel dimensions
$256\!\rightarrow\!128\!\rightarrow\!64$, kernel size $4$, stride $2$, and
padding $1$ to recover the original spatial resolution. A final convolutional layer with sigmoid activation produces the reconstructed image $\hat{\mathbf{s}}$. For $n=16$, the reconstruction decoder has 17.52M parameters.

The system is trained end-to-end using a multi-task loss function. The semantic loss is defined as the cross-entropy between the true label $\ell$ and the predicted class probabilities,
\begin{equation}
\mathcal{L}_{\text{cls}} = \mathbb{E}\left[-\log p_{\ell}(\mathbf{y}_{\text{D}})\right].
\end{equation}

The per-sample reconstruction mean squared error (MSE) is defined as
\begin{equation}
\operatorname{MSE}(\mathbf{s},\hat{\mathbf{s}})
=
\frac{
\left\|
\mathbf{s}-\hat{\mathbf{s}}
\right\|^2
}{
C N_h N_w
},
\end{equation}
where the MSE is averaged over all image elements. The reconstruction loss is then given by
\begin{equation}
\mathcal{L}_{\mathrm{rec}}
=
\mathbb{E}\left[
\operatorname{MSE}(\mathbf{s},\hat{\mathbf{s}})
\right].
\end{equation}

This loss is complemented by a structural similarity loss that captures perceptual quality. The structural similarity loss is based on the Structural Similarity Index Measure (SSIM). For input $\mathbf{s}$ and reconstruction $\hat{\mathbf{s}}$, $\text{SSIM}(\mathbf{s}, \hat{\mathbf{s}})$ evaluates the similarity between the two images by jointly comparing their local luminance, contrast, and structural information. A higher SSIM value indicates greater perceptual similarity. The corresponding loss is defined as
\begin{equation}
\mathcal{L}_{\text{ssim}} = \mathbb{E}\left[ 1 - \text{SSIM}(\mathbf{s}, \hat{\mathbf{s}}) \right].
\end{equation}

Peak signal-to-noise ratio (PSNR) is incorporated as a complementary reconstruction metric and is defined as
\begin{equation}
\operatorname{PSNR}(\mathbf{s},\hat{\mathbf{s}})
=
10\log_{10}\left(
\frac{I_{\max}^{2}}
{\operatorname{MSE}(\mathbf{s},\hat{\mathbf{s}})}
\right),
\end{equation}
where $I_{\max}=1$ denotes the maximum possible image intensity. Although derived from MSE, PSNR expresses reconstruction error on a logarithmic decibel scale. The corresponding loss term is defined as
\begin{equation}
\mathcal{L}_{\text{psnr}} = \mathbb{E}\left[ -\,\text{PSNR}(\mathbf{s}, \hat{\mathbf{s}}) \right].
\end{equation}

The total loss is formed as a weighted combination of losses,
\begin{equation}
\mathcal{L} = 
\lambda_{\text{cls}} \mathcal{L}_{\text{cls}}
+ \lambda_{\text{rec}} \mathcal{L}_{\text{rec}} 
+ \lambda_{\text{ssim}} \mathcal{L}_{\text{ssim}} 
+ \lambda_{\text{psnr}} \mathcal{L}_{\text{psnr}},
\end{equation}
where all weights are non-negative and control the trade-off between semantic accuracy, reconstruction fidelity, and perceptual quality. In the numerical results, we set $\lambda_{\text{rec}} = 5$, $\lambda_{\text{ssim}} = 2$, $\lambda_{\text{psnr}} = 0.05$, and $\lambda_{\text{cls}} = 1$.

Training is performed using the Adam optimizer with a learning rate of $10^{-3}$, a batch size of 128, and a total of 500 training epochs. During each iteration, mini-batches of input samples are encoded, transmitted through stochastic channel realizations, processed by the relay, and decoded at the destination. Gradients are computed with respect to the total loss and propagated through the network, enabling joint optimization of the source encoder, relay transformation, and destination decoders. This end-to-end framework adapts representations and processing to both channel and task requirements.

For comparison, we consider a baseline where communication occurs directly between the source and destination. The source transmits the representation $\mathbf{x}$ in a single phase without a relay, and the destination applies reconstruction and semantic decoders with the same architecture, with input dimensions adjusted to the single observation.

\begin{figure}[t!]
  \centering
    \includegraphics[width=0.95\linewidth]{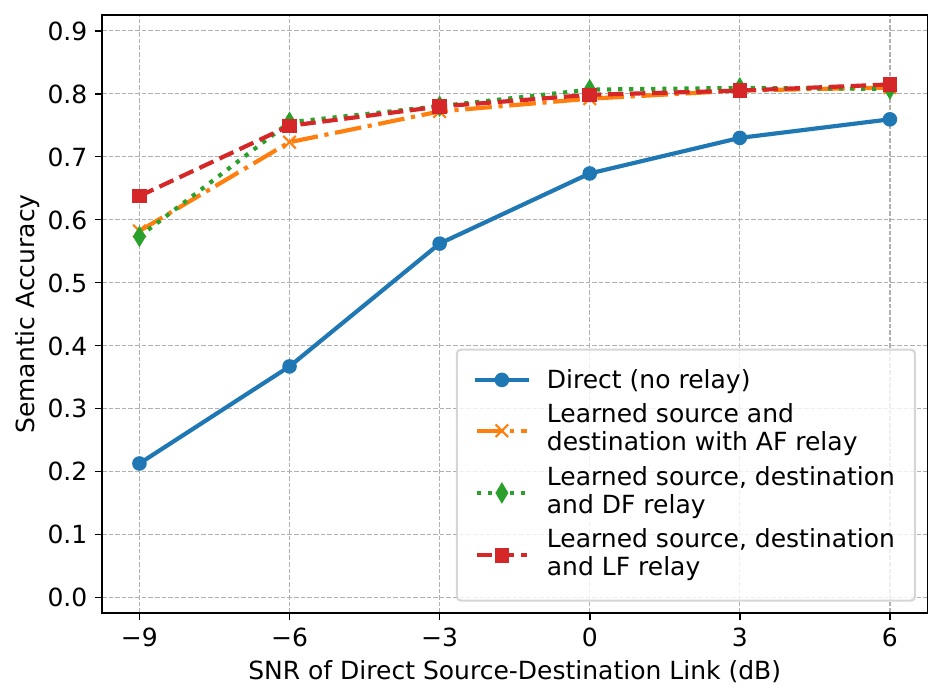}
    \caption{Semantic accuracy in semantic relay communications.}\label{fig:accuracy}
\end{figure}

For numerical results, we consider CIFAR-10 images, locate the relay midway between the source and destination, and use distance-dependent path loss with exponent 4 to determine the link SNRs. Fig.~\ref{fig:accuracy} shows the semantic accuracy as a function of the direct link SNR. Relay-assisted schemes consistently outperform the direct baseline, with gains most pronounced at low SNR. The proposed approach achieves an absolute improvement of up to $0.4248$ in semantic accuracy, along with gains of $2.58$~dB in PSNR and $0.1362$ in SSIM relative to the direct communication baseline, demonstrating improved semantic and perceptual performance.

\section{Relay Attack on Semantic Communications \label{sec:attack}}

\begin{figure}[t!]
  \centering
    \includegraphics[width=\linewidth]{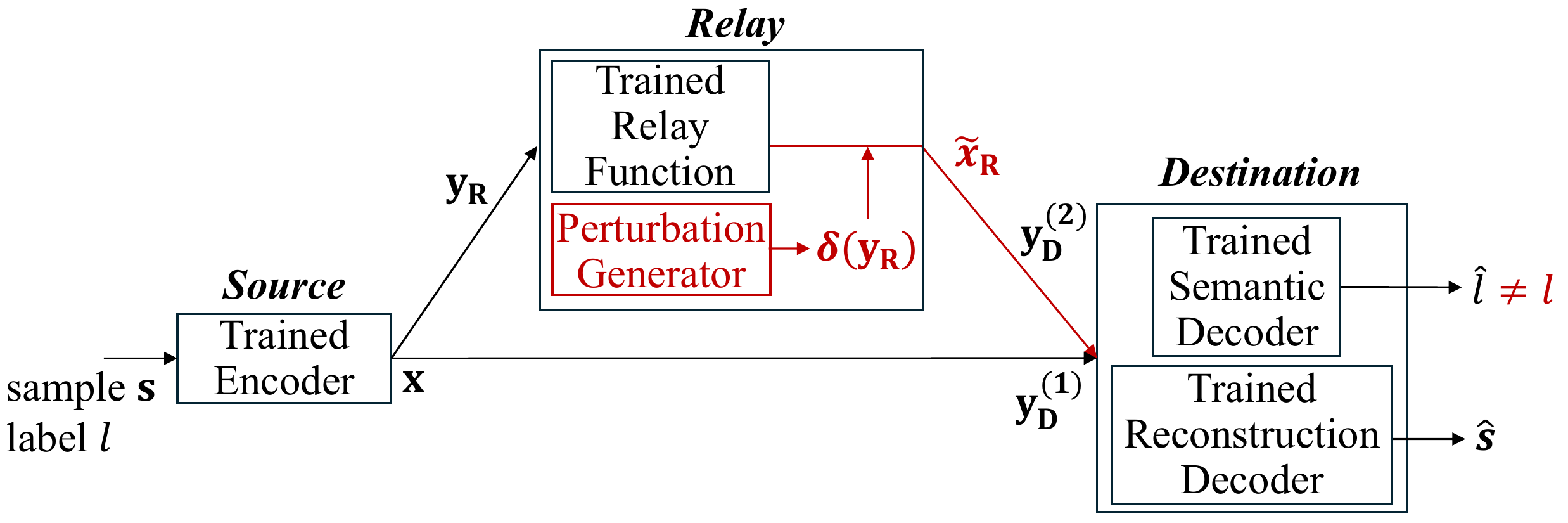}
    \caption{Attack on semantic relay communications.}\label{fig:attack}
\end{figure}

We consider a threat model in which the relay node is an insider adversary, i.e., a Byzantine node, as shown in Fig.~\ref{fig:attack}. The relay participates in the communication protocol and, as an insider in the jointly deployed system, has access to intermediate signals and knowledge of the destination processing pipeline.  The objective of the attack is to degrade the semantic inference capability of the destination while maintaining the relay transmit-power constraint.

The relay receives the signal $\mathbf{y}_{\text{R}}$ and is expected to transmit a processed version $\mathbf{x}_{\text{R}}$ toward the destination. Under normal operation, the relay applies $f_{\phi}^{R}(\cdot)$ according
to the selected relaying scheme and transmits the resulting signal. Under the Byzantine attack, the relay instead transmits a perturbed signal
\begin{equation}
\tilde{\mathbf{x}}_{\text{R}} = \mathcal{N}\left( f_{\phi}^{R}(\mathbf{y}_{\text{R}}) + \alpha \, \boldsymbol{\delta}(\mathbf{y}_{\text{R}}) \right),
\end{equation}
where $\boldsymbol{\delta}(\cdot)$ is a trained perturbation function and $\alpha > 0$ is a design parameter that controls the relative strength of the perturbation prior to normalization. The normalization $\mathcal{N}(\cdot)$ enforces unit average squared magnitude per real I/Q component, so $\alpha$ determines the relative contribution of the perturbation to the relay signal rather than the absolute transmit power. In the numerical results, $\alpha$ is set to $0.2$.

The source encoder, relay transformation, and destination decoders are jointly trained to optimize end-to-end communication performance. During attack training, the relay is assumed to have access to the architecture and parameters of the destination semantic decoder and uses them to compute the clean and attacked class probabilities. The perturbation generator is optimized offline using gradient-based training, with its parameters updated after each communication-model epoch and once more before final evaluation, while the communication model is held fixed during each update. Once trained, it generates each test-time perturbation in a single forward pass based on the current relay observation $\mathbf{y}_{\mathrm{R}}$, without access to ground-truth labels, per-sample gradient computation, or iterative test-time optimization.

The perturbation function $\boldsymbol{\delta}(\cdot)$ is implemented as a lightweight DNN operating on the relay input. The complex-valued signal $\mathbf{y}_{\mathrm{R}}\in\mathbb{C}^{n}$ is represented by its I/Q components as a real-valued tensor of size $2\times n$. The DNN consists of three one-dimensional convolutional layers with channel dimensions $2\!\rightarrow\!128\!\rightarrow\!128\!\rightarrow\!2$, kernel size $3$, and padding $1$, with ReLU activations after the first two layers. These layers capture local dependencies across consecutive channel uses, while the final layer produces a two-channel perturbation matching the relay signal dimensionality. The perturbation output is normalized to unit root-mean-square magnitude and then scaled according to the relay-input magnitude using a trainable gain. The attack DNN has $51$K trainable parameters.

The perturbation network is trained by optimizing an objective that directly
degrades semantic inference at the destination while enforcing communication
and power constraints. Let $p_k(\mathbf{y}_{\mathrm{D}})$ denote the semantic decoder's predicted
probability for class $k$ given destination observation
$\mathbf{y}_{\mathrm{D}}$, and let
$\mathbf{y}_{\mathrm{D}}^{\mathrm{clean}}$ and
$\mathbf{y}_{\mathrm{D}}^{\mathrm{adv}}$ denote the destination observations
under the clean and attacked relay transmissions, respectively.

The untargeted attack encourages prediction uncertainty by maximizing the
entropy
\begin{equation}
\mathcal{L}_{\text{ent}}
=
-\mathbb{E}\left[
\sum_{k=1}^{K}
p_k(\mathbf{y}_{\mathrm{D}}^{\mathrm{adv}})
\log p_k(\mathbf{y}_{\mathrm{D}}^{\mathrm{adv}})
\right].
\end{equation}
Maximizing this term drives the predicted
distribution toward uniformity and degrades semantic reliability.

A Kullback--Leibler (KL) divergence term encourages the attacked prediction
to differ from the clean prediction of the fixed communication model. The KL
divergence is
\begin{equation}
\mathcal{L}_{\text{kl}}
=
\mathbb{E}\left[
\sum_{k=1}^{K}
p_k(\mathbf{y}_{\mathrm{D}}^{\mathrm{clean}})
\log
\frac{
p_k(\mathbf{y}_{\mathrm{D}}^{\mathrm{clean}})
}{
p_k(\mathbf{y}_{\mathrm{D}}^{\mathrm{adv}})
}
\right].
\end{equation}
Maximizing this term drives the attacked prediction away from the clean
prediction.

To discourage large perturbations, a power regularization term is imposed,
\begin{equation}
\mathcal{L}_{\mathrm{pow}}
=
\mathbb{E}\left[
\frac{
\left\|
\boldsymbol{\delta}(\mathbf{y}_{\mathrm{R}})
\right\|^2
}{
2n
}
\right],
\end{equation}
which penalizes the mean squared perturbation magnitude and controls the perturbation magnitude prior to
the normalization step that enforces the relay transmit-power constraint.

The overall attack objective is defined as
\begin{equation}
\mathcal{L}_{\mathrm{attack}}
=
- \lambda_{\mathrm{ent}} \mathcal{L}_{\mathrm{ent}}
- \lambda_{\mathrm{kl}} \mathcal{L}_{\mathrm{kl}}
+ \lambda_{\mathrm{pow}} \mathcal{L}_{\mathrm{pow}}.
\end{equation}
Minimizing $\mathcal{L}_{\mathrm{attack}}$ maximizes $\mathcal{L}_{\mathrm{ent}}$ to increase prediction uncertainty and $\mathcal{L}_{\mathrm{kl}}$ to separate the attacked and clean predictions, while minimizing $\mathcal{L}_{\mathrm{pow}}$ to penalize perturbation power. In the numerical results, we set
$\lambda_{\mathrm{ent}}=1$ and
$\lambda_{\mathrm{kl}}=1$, and vary
$\lambda_{\mathrm{pow}}$. The perturbation network is trained separately using stochastic gradient-based optimization while the communication-model parameters are frozen, with only the perturbation-network parameters updated. At test time, the perturbation is generated in a single forward pass without access to ground-truth labels.

Unlike standard input-space gradient-based attacks such as the Fast Gradient Sign Method (FGSM) and Projected Gradient Descent (PGD), the proposed attack operates on latent communication signals and uses a learned mapping to generate perturbations in a single forward pass at test time, without per-sample gradient computation or optimization. It is constrained by physical-layer effects, including power normalization and channel impairments, and does not require label access at test time. Rather than manipulating decoder inputs directly, the relay perturbs an intermediate representation that propagates through the wireless channel. The perturbation model is trained over varying channel realizations.

\begin{figure}[t!]
  \centering
    \includegraphics[width=0.95\linewidth]{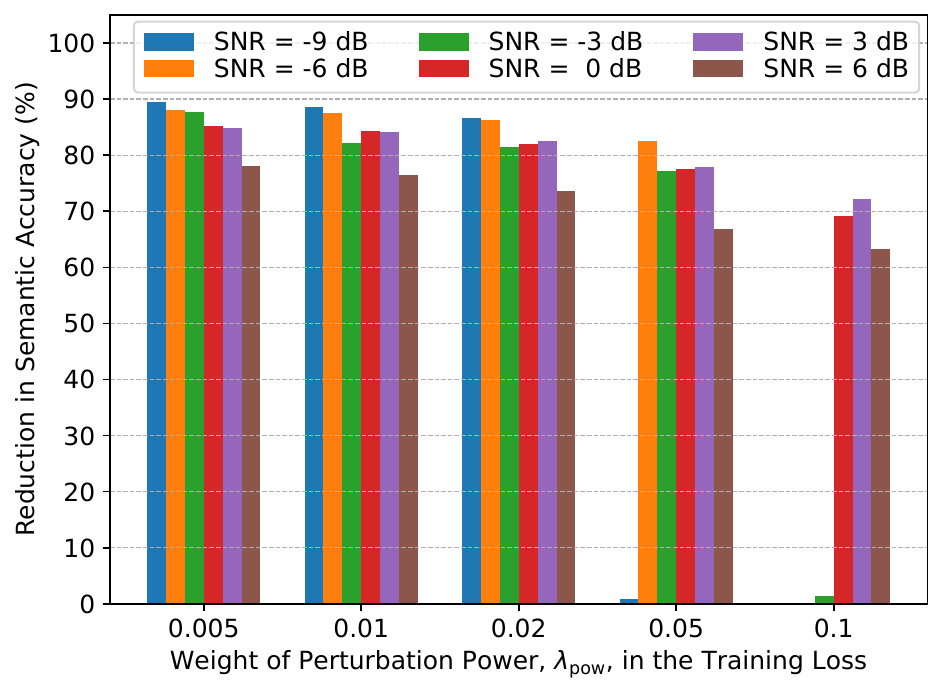}
    \caption{Reduction in semantic accuracy due to relay attack.}\label{fig:attack_effect}
\end{figure}

Fig.~\ref{fig:attack_effect} shows the reduction in semantic accuracy due to the relay attack under the LF strategy. The attack is highly effective across the considered SNR range, causing major degradation in semantic accuracy as the direct source-destination SNR varies. Its impact is governed by the perturbation power weight $\lambda_{\text{pow}}$, which controls the trade-off between attack strength and perturbation magnitude. As $\lambda_{\text{pow}}$ increases, the perturbation is more constrained and the attack effect is reduced, primarily at very low SNR where channel noise dominates, but becomes prominent again at moderate and high SNR even under stronger constraints. Across all considered SNRs and values of $\lambda_{\text{pow}}$, the attack
reduces semantic accuracy by up to $89.42\%$ relative to the no-attack case and degrades perceptual quality, with PSNR and SSIM reductions of up to
$5.75$~dB and $0.2317$, respectively, potentially driving performance below
the direct transmission baseline. The attack remains effective, with lower $\lambda_{\text{pow}}$ amplifying its impact across a wider SNR range.

\section{Conclusion \label{sec:conclusion}}
We have studied a two-phase relay-assisted SemCom model, where an intermediate relay processes received wireless signals to improve end-to-end semantic accuracy. Joint optimization of the source encoder, relay processing, and destination decoders improves semantic accuracy over direct transmission, with larger gains at lower SNRs. To examine the associated security risks, we considered the relay as an insider (Byzantine) adversary that exploits access to intermediate signals and knowledge of the destination model to inject structured perturbations. We have shown that a perturbation function trained via deep learning can effectively degrade semantic inference at test time. These results highlight the major security risks introduced by relay-assisted SemCom despite their semantic-accuracy gains, motivating robust defenses against insider threats in learning-enabled relay systems.
\bibliographystyle{IEEEtran}
\bibliography{references}

\end{document}